\documentstyle[11pt,paspconf,natbib]{article}
\bibpunct[,]{(}{)}{,}{a}{}{,}
\makeatletter
\def\thebibliography#1{{\section*{\refname}\@mkboth
  {\refname}{\refname}}
  \typeout{\refname}\def\bibwidthlabel{#1}\list
  {\kapbib@counter}{\kapbib@list}
    \let\makelabel\@biblabel
    \def\newblock{\hskip .11em plus .33em minus .07em}
    \sloppy\clubpenalty4000\widowpenalty4000
    \sfcode`\.=1000\relax}

  \def\kapbib@counter{\relax}
  \def\kapbib@list{\setlength{\labelsep}{0em}%
        \setlength{\labelwidth}{0pt}%
        \setlength{\itemindent}{-\bibhang}%
        \setlength{\itemsep}{0pt}%
        \setlength{\parsep}{0.5ex}
        \setlength{\leftmargin}{\bibhang}}
  \def\@biblabel#1{}
\makeatother
\makeatletter
\def\mpspacing#1{\def\baselinestretch{#1}\let\glb@currsize=\relax\selectfont}
\renewcommand{\fps@figure}{thbp}
\renewcommand{\fps@table}{thbp}

\makeatother
\def\tablefontsize{\small\rm}

\def\mag{\hbox{$.\!\!^{\rm m}$}}

\def\la{\mathrel{\hbox{\rlap{\hbox{\lower4pt\hbox{$\sim$}}}\hbox{$<$}}}}
\def\ga{\mathrel{\hbox{\rlap{\hbox{\lower4pt\hbox{$\sim$}}}\hbox{$>$}}}}
\def\kms{\nobreak\mbox{$\;$km\,s$^{-1}$}}

\begin{document}
%

\title{The Luminosity Function of Globular Clusters as
       an Extragalactic Distance Indicator}

\author{G.A. TAMMANN}

\affil{Astronomisches Institut der Universit{\"a}t Basel, \\
       Venusstr.~7, CH-4102 Binningen, Switzerland}

\author{A. SANDAGE}

\affil{The Observatories of the Carnegie Institution of Washington, \\ 
         813 Santa Barbara Street, Pasadena, CA 91191--1292, USA}

\begin{abstract}
  The absolute magnitude $M^{\ast}$ of the peak of the globular
  cluster luminosity function (GCLF), approximated by a Gaussian, can
  be calibrated independently in the Galaxy and M\,31 through RR\,Lyr
  stars and Cepheids, respectively. They yield, in perfect agreement,
  $M^{\ast}_{\rm B} = -6.93\pm0.08$ and  $M^{\ast}_{\rm V} =
  -7.62\pm0.08$. Application of these values to the GCLFs in the Leo
  Group $(n=2)$, the Virgo cluster $(n=8)$, and the Coma cluster
  $(n=2)$ gives distances which agree with the best determinations
  from other methods. However, the corresponding distance of the
  Fornax cluster $(n=7)$ is significantly underestimated, and the
  distances of several field galaxies are inconsistent. A second
  parameter, like the width of the GCLF or the color of the peak, is
  apparently needed to control differences in the GC formation
  history. 
\end{abstract}

\section{Introduction}
\label{sec:introduction}
Extragalactic GCs, discovered in M\,31 by \citet{Hubble:32}, took a
role in distance determinations when \citet{Racine:68} proposed the
bright end of the GCLF to be used as a ``standard candle''. First
applications of this tool provided reasonable distances to M\,87
\citep{Sandage:68,deVaucouleurs:70}, yet it was soon realized that the
results were sensitive to the GC population size, and that the
luminosity $M^{\ast}$ of the turnover point of the bell-shaped GCLF is
a much more stable standard candle. This required, however, that one
had to sample at least four magnitudes into the GCLF which became
feasible only with the advent of CCDs. The first application of the
new method to a giant E galaxy \citep*[M\,87;][]{vandenBergh:etal:85}
was followed by many papers such that $M^{\ast}$ magnitudes are now
available for about two dozen full-size galaxies 
\citep[for reviews e.g.][]{Harris:91,Whitmore:97}. 

   Of course distances from the turnover of the GCLF require also an
absolute, local calibration of $M^{\ast}$. This became available as
sufficient data of complete and objective samples of the GCs in the
Galaxy and M\,31 accumulated which could be combined with a reliable
calibration of RR\,Lyr star luminosities and the Cepheid distance of
M\,31, respectively \citep{Sandage:Tammann:95}.

   The calibration of $M^{\ast}$ is discussed in Section~2. The
resulting distances of 23 E~galaxies and one spiral are compared
with external evidence in Section~3, the main purpose being to test
the stability of $M^{\ast}$ as a standard candle. The conclusions
are in Section~4.

\section{The calibration of the GCLF}
\label{sec:2}

\subsection{The Galactic GCLF}
\label{sec:2_1}
A list of the 100 Galactic GCs, which an external observer would
observe, has been compiled by \citet{Secker:92}. Their $B,V$
magnitudes, horizontal-branch magnitudes, extinction values, and
metallicities have been compiled by \citet[Table~1]{Sandage:Tammann:95};
the individual sources are given there. The
absolute magnitudes $M^{\ast}_{\rm B}$ and $M^{\ast}_{\rm V}$ are
calculated from Sandage's (\citeyear{Sandage:93c}) RR\,Lyr star
calibration
\begin{equation}\label{equ:1}
  M_{\rm V}({\rm RR}) = 0.94 + 0.30 ([{\rm Fe/H}]),
\end{equation}
which is based on the observed position of the blue edge of the
RR\,Lyr instability strip and the requirements of pulsation theory. 

   The peak of a Gaussian LF, calculated from equation~(\ref{equ:1}),
depends only on the {\em mean\/} metallicity of the sample, which is
in this case $<\![{\rm Fe/H}]\!>=-1.35$. Changes of the slope of the
metallicity term in equation~(\ref{equ:1}) affects only the width of
the LF to some extent. Calibrations of RR\,Lyr stars agree amazingly
well at a metallicity of $[{\rm Fe/H}]=-1.35$ as seen in
Table~\ref{tab:1}. 
The Galactic calibration involving proper motions tend to give
somewhat fainter RR\,Lyr luminosities
\citep{Fernley:etal:98,Tsujimoto:98}, but the results still agree with
the adopted value of $M_{\rm V}({\rm RR},-1.35)=0.54$ to within 
$\la 1$ sigma.
%
\begin{table}[hb]
\tablefontsize
\begin{center}
\caption{$M_{\rm V}({\rm RR})$ at $[{\rm Fe/H}]=-1.35$ from various authors}
\label{tab:1}
\begin{tabular}{lrl}
\noalign{\smallskip}
\hline
\noalign{\smallskip}
  blue edge $+$ pulsation theory & 0.54 & \citealt{Sandage:93c} \\ 
  Baade-Becker-Wesselink         & 0.55 & \citealt{Cacciari:98} \\  
  Hipparcos                      & 0.54 & \citealt{Reid:98} \\         
  Hipparcos                      & 0.48 & \citealt{Gratton:etal:97} \\
  review                         & 0.52 & \citealt{Chaboyer:etal:98} \\ 
\noalign{\smallskip}
\hline
\noalign{\smallskip}
  adopted                        & 0.54 & \\
\tableline
\end{tabular}
\end{center}
\end{table}

   The calibration of \citet{Sandage:93c} at the specific metallicity
seems secure within a few $0\mag01$. The corresponding peak magnitudes
of the Galactic GCLF are therefore repeated here from
\citet{Sandage:Tammann:95} 
\begin{eqnarray}\label{equ:2}
  M^{\ast}_{\rm B}({\rm Galaxy}) & = & -6.90\pm0.11, \qquad
         \sigma_{\rm M}=1.07\pm0.11 \\
\label{equ:3}
  M^{\ast}_{\rm V}({\rm Galaxy}) & = & -7.60\pm0.11, \qquad
         \sigma_{\rm M}=1.07\pm0.11.
\end{eqnarray}

\subsection{The GCLF of M\,31}
\label{sec:2_2}
The LF of Secker's (\citeyear{Secker:92}) unbiased sample of 82 GCs in
M\,31 peaks at apparent magnitudes of $m_{\rm B}^{\ast}=17.75\pm0.11$
and $m_{\rm V}^{\ast}=16.98\pm0.11$ \citep{Sandage:Tammann:95}. This
translates into
\begin{eqnarray}\label{equ:4}
  M^{\ast}_{\rm B}({\rm M\,31}) & = & -7.01\pm0.20, \qquad
         \sigma_{\rm M}=0.89\pm0.10 \\
\label{equ:5}
  M^{\ast}_{\rm V}({\rm M\,31}) & = & -7.70\pm0.20, \qquad
         \sigma_{\rm M}=0.89\pm0.10 
\end{eqnarray}
with a true M\,31 modulus of $(m-M)^{0}=24.44\pm0.15$
\citep{Madore:Freedman:91} and a mean reddening of the GCs of
$E(B-V)=0.08$ \citep{Sandage:Tammann:95}.

   A purist may want to base the GC distance scale on Pop~II stars
exclusively instead of involving Cepheids. In that case it may be
noted that according to \citet{Pritchet:vandenBergh:87a} the RR\,Lyr
stars in M\,31 have $<\!\!m_{\rm B}\!\!>_{\rm RR}=25.68\pm0.06$,
which gives with $<\!\!B-V\!\!>_{\rm RR}=0.26$ \citep{Hawley:etal:86}
$<\!\!m_{\rm V}\!\!>=25.42\pm0.06$.
Adopting $[{\rm Fe/H}]=-0.6$ from Pop~II giants in M\,31
\citep{Mould:Kristian:86} implies $M_{\rm V}({\rm RR})=0.76$ from
equation~(\ref{equ:1}).
The resulting apparent M\,31 modulus of $(m-M)_{\rm AV}=24.66$ is the
same as that from Cepheids to within $0\mag02$. Alternatively
\citet{Holland:98} finds from fitting the red-giant branches of M\,31
GCs a true modulus of $(m-M)^{0}=24.47\pm0.07$ which is only
insignificantly larger than the Cepheid distance.

\subsection{The combined GCLF of the Galaxy and M\,31}
\label{sec:2_3}
Noting the perfect agreement of the independent calibrations of
equations~(\ref{equ:2}), (\ref{equ:3}) and (\ref{equ:4}), (\ref{equ:5}) we
have made a joint Gaussian fit to the absolute magnitudes of the GCs
in the Galaxy and M\,31, resulting in 
\begin{eqnarray}\label{equ:6}
  M^{\ast}_{\rm B}({\rm adopted}) & = & -6.93\pm0.08, \qquad
         \sigma_{\rm M}=1.02\pm0.08 \\
\label{equ:7}
  M^{\ast}_{\rm V}({\rm adopted}) & = & -7.62\pm0.08, \qquad
         \sigma_{\rm M}=1.02\pm0.08 
\end{eqnarray}
\citep[cf.][]{Sandage:Tammann:95,Sandage:Tammann:96a}.
\section{Applications of the calibrated GCLF}
\label{sec:3}

\subsection{The Leo Group}
\label{sec:3_1}
For two galaxies of the Leo group \citep*[cf.][]{Humason:etal:56} 
the turnover point $m^{\ast}$ of the GCLF has been determined 
(Table~\ref{tab:2}). The assigned errors are our estimates. 
The resulting distance is based on equation~(\ref{equ:6}).
%
\begin{table}[hb]
\tablefontsize
\begin{center}
\caption{$m^{\ast}$ values of the Leo group}
\label{tab:2}
\begin{tabular}{lrl}
\noalign{\smallskip}
\hline
\noalign{\smallskip}
  galaxy       & \multicolumn{1}{c}{$m^{\ast}_{\rm B}$} & Source \\
\noalign{\smallskip}
\hline
\noalign{\smallskip}
  NGC\,3377/79 & $23.30 \pm 0.40$ & \citealt{Pritchet:vandenBergh:85} \\
  NGC\,3379    & $23.00 \pm 0.40$ & \citealt{Harris:90} \\    
\noalign{\smallskip}
\hline
\noalign{\smallskip}
  mean:        & $23.15 \pm 0.28$ & \\
  \multicolumn{2}{r}{$\Rightarrow\;\; (m-M)=30.08\pm0.29$} &  (10.4$\;$Mpc)\\
\noalign{\smallskip}
\hline
\noalign{\smallskip}
\end{tabular}
\end{center}
\end{table}

   The best distance of the Leo group is obtained from the three
members whose Cepheids have been observed with {\sl HST\/}
(Table~\ref{tab:3}). The moduli are corrected by $+0\mag05$ for the
long/short exposure effect of the WFPC2 \citep[cf.][]{Saha:etal:96a}.  

%
\begin{table}[hb]
\tablefontsize
\begin{center}
\caption{The Cepheid distances of the Leo group}
\label{tab:3}
\begin{tabular}{lrl}
\noalign{\smallskip}
\hline
\noalign{\smallskip}
  galaxy       & \multicolumn{1}{c}{$(m-M)$}  & Source \\
\noalign{\smallskip}
\hline
\noalign{\smallskip}
  NGC\,3351 & $30.01 \pm 0.19$ & \citealt{Graham:etal:97} \\
  NGC\,3368 & $30.37 \pm 0.16$ & \citealt{Tanvir:etal:95} \\    
  NGC\,3627 & $30.22 \pm 0.22$ & \citealt{Saha:etal:98a} \\    
\noalign{\smallskip}
\hline
\noalign{\smallskip}
  mean:        & $30.20 \pm 0.12$ & (10.9$\;$Mpc) \\
\noalign{\smallskip}
\hline
\noalign{\smallskip}
\end{tabular}
\end{center}
\end{table}

   The  distance modulus difference of $0.10\pm0.35$ between
Tables~\ref{tab:2} and \ref{tab:3} is fortuitously small.

\subsection{The Virgo cluster}
\label{sec:3_2}
Different values of $m^{\ast}_{\rm B}$ and $m^{\ast}_{\rm V}$ of bona fide
members of the Virgo cluster \citep*[cf.][]{Binggeli:etal:85} are
compiled in Table~\ref{tab:4}. The values have been corrected for the
small Galactic absorption according to \citet{Burstein:Heiles:84}. The
$g$ magnitudes of \citet{Cohen:88} and the $R$ magnitudes of
\citet{Ajhar:etal:94} were transformed into $V$ magnitudes following
\citet{Whitmore:97}. No (precarious) attempt was made to correct
$m^{\ast}_{\rm B}$ into $m^{\ast}_{\rm V}$. 

%
\begin{table}[hbt]
\tablefontsize
\begin{center}
\caption{$m^{\ast}$ values of the Virgo cluster}
\label{tab:4}
\begin{minipage}{0.9\textwidth}
\begin{tabular}{lllrc}
\noalign{\smallskip}
\hline
\noalign{\smallskip}
  Galaxy & \multicolumn{1}{c}{$m^{\ast}_{\rm B}$} &
  \multicolumn{1}{c}{$m^{\ast}_{\rm V}$}   & 
  \multicolumn{1}{c}{$m^{\ast}_{\rm B}$ - $m^{\ast}_{\rm V}$}  & Source \\
\noalign{\smallskip}
\hline
\noalign{\smallskip}
 NGC\,4365 & $25.18\pm0.16(2)$ & $24.47\pm0.21(1)$ & $0.71\pm0.26$ & (1)\\
 NGC\,4374 &                   & $24.12\pm0.30(1)$ &               & (2)\\
 NGC\,4406 &                   & $24.25\pm0.30(1)$ &               & (2)\\
 NGC\,4472 & $24.70\pm0.11(1)$ & $23.85\pm0.21(2)$ & $0.85\pm0.24$ & (3)\\
 NGC\,4486 & $24.82\pm0.11(2)$ & $23.74\pm0.06(5)$ & $1.08\pm0.13$ & (4)\\
 NGC\,4552 &                   & $23.70\pm0.30(1)$ &               & (2)\\
 NGC\,4636 &                   & $24.18\pm0.20(1)$ &               & (5)\\
 NGC\,4649 & $24.65\pm0.14(1)$ &                   &               & (6)\\ 
\noalign{\smallskip}
\hline
\noalign{\smallskip}
 straight mean: & $24.84\pm0.12$ & $24.03\pm0.10$ & & \\
 $(m-M)$:       & $31.77\pm0.14$ & $31.65\pm0.13$ & & \\
\noalign{\smallskip}
 $\Rightarrow$  & \multicolumn{2}{l}{$(m-M)=31.70\pm0.10$} & (21.9$\;$Mpc)\\ 
\noalign{\smallskip}
\hline
\noalign{\smallskip}
\end{tabular}
\mpspacing{0.8}

{\footnotesize
  Sources: 
 (1) \citealt{Harris:etal:91}; 
     \citealt{Secker:Harris:93};
     \citealt{Forbes:96a}
 (2) \citealt{Ajhar:etal:94}
 (3) \citealt{Harris:etal:91};
     \citealt{Ajhar:etal:94};
     \citealt{Cohen:88}
 (4) \citealt{vandenBergh:etal:85};
     \citealt{Harris:etal:91};
     \citealt{Cohen:88};
     \citealt{McLaughlin:etal:94};
     \citealt{Whitmore:etal:95};
     \citealt{Elson:Santiago:96a,Elson:Santiago:96b} 
 (5) \citealt{Kissler:etal:94}
 (6) \citealt{Harris:etal:91}.
  -- The values in parentheses in columns~2 and 3 give the number of
  independent determinations.}
\end{minipage}
\end{center}
\end{table}

   The distance moduli in $B$ and $V$ in Table~\ref{tab:4} follow from
equation (\ref{equ:6}) and (\ref{equ:7}). The former is marginally
larger than the latter. The weighted mean GCLF modulus of the Virgo
cluster is $(m-M)=31.70\pm0.10$.

   Independent distance determinations of the Virgo cluster are as
   follows: 

Cepheids at the distance of the Virgo cluster are difficult objects
even for {\sl HST\/}. There are now three bona fide cluster members 
and two outlying members with Cepheid distances from {\sl HST\/}.
The wide range of their distance moduli, corresponding to $14.9$ to
$25.5\;$Mpc, reveals the important depth effect of the cluster. Four
of the galaxies  have been chosen from the
atlas of \citet{Sandage:Bedke:88} because they are highly resolved and
seemed easy as to their Cepheids. They are therefore {\em expected\/}
to lie on the near side of the cluster. In contrast NGC\,4639 has been
chosen as parent to SN\,1990N and hence independently of its distance;
correspondingly this distance is expected to be statistically more
representative. A straight mean of the distances is
therefore likely to be an underestimate. Indeed the mean Tully-Fisher
(TF) distance modulus of the five galaxies is $0\mag2$ (corresponding to
$10\%$ in distance) {\em smaller\/} than the mean cluster distance of a
complete and fair sample of TF distances \citep{Federspiel:etal:98}.

   A preliminary Cepheid distance of the Virgo cluster is obtained by
taking the Cepheid distance of the Leo group of $(m-M)=30.20\pm0.12$,
based now on three galaxies with Cepheids from {\sl HST}, and to step
up this value by the modulus difference of $\Delta(m-M)=1.25\pm0.13$
\citep{Tammann:Federspiel:97} between the Leo group and the Virgo
cluster. The corresponding result is shown in Table~\ref{tab:5}.
%
\begin{table}[hb]
\tablefontsize
\begin{center}
\caption{The Virgo cluster modulus from various methods excluding the GCLF}
\label{tab:5}
\begin{tabular}{lllc}
\noalign{\smallskip}
\hline
\noalign{\smallskip}
        Method               & $(m-M)_{\rm Virgo}$ & Hubble type & Source\\
\noalign{\smallskip}
\hline
\noalign{\smallskip}
        Cepheids              &  $31.45 \pm  0.21$ &     S       & 1 \\
        Tully-Fisher          &  $31.58 \pm  0.24$ &     S       & 2 \\
        SNe\,Ia               &  $31.52 \pm  0.20$ &     E, S    & 3 \\
        D$_{\rm n} - \sigma$  &  $31.85 \pm  0.19$ &     S0, S   & 4 \\
        Novae                 &  $31.46 \pm  0.40$ &     E       & 5 \\
\noalign{\smallskip}
\hline
\noalign{\smallskip}
        Mean:                 &  $31.61 \pm  0.09$ &
 ($\Rightarrow 21.0\pm0.9\;$Mpc) \\
\noalign{\smallskip}
\hline
\noalign{\smallskip}
\end{tabular}
\begin{minipage}{0.9\textwidth}\footnotesize
Sources:
\begin{enumerate}
\def\parsep{0pt}
\def\itemsep{0pt}
\item See text.
\item \citet{Federspiel:etal:98}.
\item See text.
\item A reasonably tight D$_{\rm n}-\sigma$ relation of S0 and spiral
  galaxies in the Virgo cluster has been published by
  \citet{Dressler:87}. The zeropoint calibration rests on the distance
  of the Galactic bulge ($7.8\;$kpc) and the Cepheid distances of
  M\,31 and M\,81 \citep{Sandage:Tammann:88,Tammann:88}.
\item \citet{Pritchet:vandenBergh:87} found from six novae in Virgo
  cluster ellipticals that they are $7\mag0\pm0\mag4$ more
  distant than the {\em apparent\/} distance modulus of M\,31 of
  $(m-M)_{\rm AB}=24.58\pm0.10$ from Cepheids
  \citep{Madore:Freedman:91} {\em and\/} Galactic novae
  \citep{Capaccioli:etal:89}. \citet{Livio:97} found from a
  semi-theoretical analysis of the six Virgo novae $(m-M)_{\rm
  Virgo}=31.35\pm0.35$. A low-weight mean of $31.46$ is adopted.
\end{enumerate}
\end{minipage}
\end{center}
\end{table}

   A good Virgo cluster distance comes from the relation between the
21cm line width of inclined spirals and their luminosity (the
so-called Tully-Fisher relation). Because the method, calibrated by 20
local galaxies with Cepheid distances, can be applied to a {\em
  complete\/} sample of Virgo spirals the result is exceptionally
immune to selection effects which systematically lead to an
underestimate of all distances (the ever-present Malmquist effect of
magnitude-limited samples). The resulting distance modulus is shown in
Table~\ref{tab:5}. 

A high-weight Virgo modulus comes also from the seven SNe\,Ia
(excluding the peculiar SN\,1991T) which give $<\!m_{\rm
  B}\!>\,=12.01\pm0.16$ and $<\!m_{\rm  V}\!>\,=11.96\pm0.17$ at maximum. The
absolute magnitude of SNe\,Ia is well calibrated from eight SNe\,Ia
with Cepheid distances \citep{Saha:etal:98a}. From this the Virgo
modulus becomes $(m-M)=31.46\pm0.17$.
Four of the seven SNe\,Ia have known $\Delta m_{15}$ values. If they
are corrected for $\delta\Delta m_{15}$ and $\delta (B-V)$
\citep[cf.][]{Parodi:etal:98}) the Virgo cluster modulus becomes
$31.52\pm0.20$ as included in Table~\ref{tab:5}. 

   Table~\ref{tab:5} lists also two additional distance
determinations of the Virgo cluster. For details the reader is
referred to the original literature.

The Virgo cluster modulus from the GCLF and from other methods
compares very well, the difference being $\Delta
(m-M)=-0\mag09\pm0.13$. 

\subsection{The Coma cluster}
\label{sec:3_3}
For two Coma cluster members $m^{\ast}_{\rm V}$ values are available from
{\sl HST\/} observations (Table~\ref{tab:6}).

With the adopted value of $m^{\ast}_{\rm V}$ and equation~(\ref{equ:7})
one obtains the Coma cluster distance as shown in Table~\ref{tab:6}.

%
\begin{table}[ht]
\tablefontsize
\begin{center}
\caption{$m^{\ast}$ values of the Coma cluster}
\label{tab:6}
\begin{tabular}{lrl}
\noalign{\smallskip}
\hline
\noalign{\smallskip}
  galaxy       & \multicolumn{1}{c}{$m^{\ast}_{\rm V}$} & Source \\
\noalign{\smallskip}
\hline
\noalign{\smallskip}
  NGC\,4481 & $>27.3 \pm 0.3$ & \citealt{Baum:etal:95} \\
  IC\,4051  & $27.72 \pm 0.3$ & \citealt{Baum:etal:97} \\    
\noalign{\smallskip}
\hline
\noalign{\smallskip}
  adopted:        & $27.72 \pm 0.3$ & \\
\noalign{\smallskip}
  \multicolumn{2}{l}{$\Rightarrow \;\;(m-M)=35.34\pm0.31$} &  (117$\;$Mpc)\\
\noalign{\smallskip}
\hline
\noalign{\smallskip}
\end{tabular}
\end{center}
\end{table}

   The modulus difference Coma-Virgo from four different methods in
Table~\ref{tab:7} together with the Virgo cluster modulus in Table~
\ref{tab:5} gives $(m-M)=35.32\pm0.13$ for the Coma cluster. The
agreement with the GCLF distance in Table~\ref{tab:6} is fortuitously
good.

%
\begin{table}[ht]
\tablefontsize
\begin{center}
\caption{Independent determinations of the Coma cluster modulus
  relative to the Virgo cluster}
\label{tab:7}
\begin{tabular}{lrl}
\noalign{\smallskip}
\hline
\noalign{\smallskip}
  Method       & $\Delta(m-M)_{\rm Coma-Virgo}$ & Source \\
\noalign{\smallskip}
\hline
\noalign{\smallskip}
  1st-ranked gal.    & $3.34\pm0.30$ & \citealt{Sandage:etal:76} \\
  10 brightest gal.  & $4.16\pm0.20$ & \citealt{Weedman:76} \\ 
  D$_{\rm n}-\sigma$ & $3.74\pm0.14$ & \citealt{Faber:etal:89} \\
                     & $3.55\pm0.15$ & \citealt{DOnofrio:etal:97} \\
  TF + D$_{\rm n}-\sigma$ & $3.70\pm0.14$ & \citealt{Dekel:95} \\
\noalign{\smallskip}
\hline
\noalign{\smallskip}
  weighted mean:        & $3.71 \pm 0.08$ & \\
\noalign{\smallskip}
 $+(m-M)_{\rm Virgo} \;\;\Rightarrow$ & $35.32\pm0.13$ & \\ 
\noalign{\smallskip}
\hline
\noalign{\smallskip}
\end{tabular}
\end{center}
\end{table}

\subsection{The Fornax cluster}
\label{sec:3_4}
For seven galaxies in the Fornax cluster $m^{\ast}_{\rm B}$ and/or 
$m^{\ast}_{\rm V}$ values are available (Table~\ref{tab:8}).

%
\begin{table}[ht]
\tablefontsize
\begin{center}
\caption{$m^{\ast}_{\rm B}$ and $m^{\ast}_{\rm V}$ values of the
  Fornax cluster} 
\label{tab:8}
\begin{minipage}{0.9\textwidth}
\begin{tabular}{lllrl}
\noalign{\smallskip}
\hline
\noalign{\smallskip}
  Galaxy & \multicolumn{1}{c}{$m^{\ast}_{\rm B}$} &
  \multicolumn{1}{c}{$m^{\ast}_{\rm V}$}   & 
  \multicolumn{1}{c}{$m^{\ast}_{\rm B}$ - $m^{\ast}_{\rm V}$}  &
  \multicolumn{1}{c}{Source} \\
\noalign{\smallskip}
\hline
\noalign{\smallskip}
 NGC\,1344 &                  & $23.28\pm0.25(1)$ &        & (1)\\
 NGC\,1374 &                  & $23.52\pm0.14(1)$ &        & (2)\\
 NGC\,1379 &                  & $23.68\pm0.28(1)$ &        & (2)\\
 NGC\,1380 & $24.38\pm0.2(1)$ & $23.86\pm0.15(2)$ & $0.48$ & (1)(3)\\
 NGC\,1399 & $24.72\pm0.2(1)$ & $23.84\pm0.10(5)$ & $0.88$ & (1)(2)(4)\\
 NGC\,1404 &                  & $23.96\pm0.15(2)$ &        & (1)(5)\\
 NGC\,1427 &                  & $23.78\pm0.21(1)$ &        & (2)\\
\noalign{\smallskip}
\hline
\noalign{\smallskip}
 straight mean: & $24.55\pm0.20$ & $23.70\pm0.10$ & $0.85\pm0.22$ & \\
\noalign{\smallskip}
 $(m-M)$:       & $31.48\pm0.20$ & $31.32\pm0.10$ & & \\
 $\Rightarrow$  & \multicolumn{2}{l}{$(m-M)=31.35\pm0.10$} & (18.6$\;$Mpc)\\ 
\noalign{\smallskip}
\hline
\noalign{\smallskip}
\end{tabular}
\mpspacing{0.8}

{\footnotesize
  Sources: 
 (1) \citealt{Blakeslee:Tonry:96} 
 (2) \citealt{Kohle:etal:96}
 (3) \citealt{DellaValle:etal:98}
 (4) \citealt{Madejsky:Bender:90};
     \citealt{Geisler:Forte:90};
     \citealt{Bridges:etal:91}
 (5) \citealt{Richtler:etal:92}.
  -- The values in parentheses in columns~2 and 3 give the number of
  independent determinations.}
\end{minipage}
\end{center}
\end{table}

   The best independent distance of the Fornax cluster is obtained
from its three SNe\,Ia (SN\,1980N, 1981D, and 1992A). They have
occurred in early-type galaxies and are therefore likely to lie in
the {\em center\/} of the cluster. With the luminosity calibration
of eight local SNe\,Ia through Cepheids
\citep{Saha:etal:98a,Macchetto:98} one obtains for them a Fornax
cluster modulus of $(m-M)=31.68\pm0.15$ after full correction for
second parameter differences \citep{Parodi:etal:98}. 

   A different route to the Fornax modulus is through distance
determinations {\em relative\/} to the Virgo cluster as compiled in
Table~\ref{tab:9}. 

%
\begin{table}[ht]
\tablefontsize
\begin{center}
\caption{Independent determinations of the Fornax cluster modulus
  relative to the Virgo cluster}
\label{tab:9}
\begin{tabular}{lrl}
\noalign{\smallskip}
\hline
\noalign{\smallskip}
  Method       & $\Delta(m-M)_{\rm Fornax-Virgo}$ & Source \\
\noalign{\smallskip}
\hline
\noalign{\smallskip}
  1st-ranked gal.    & $0.44\pm0.30$ & \citealt{Sandage:etal:76} \\
  D$_{\rm n}-\sigma$ & $0.14\pm0.16$ & \citealt{Faber:etal:89} \\
                     & $0.45\pm0.15$ & \citealt{DOnofrio:etal:97} \\
  SB of dwarfs       & $0.28\pm0.08$ & \citealt{Jerjen:Binggeli:97} \\ 
  SBF                & $0.20\pm0.08$ & \citealt{Tonry:97} \\ 
  PNe                & $0.32\pm0.10$ & \citealt{Jacoby:97} \\ 
\noalign{\smallskip}
\hline
\noalign{\smallskip}
  weighted mean:        & $0.28 \pm 0.06$ & \\
\noalign{\smallskip}
 $+(m-M)_{\rm Virgo}\;\;\Rightarrow$  & $31.89\pm0.11$ & \\ 
\noalign{\smallskip}
\hline
\noalign{\smallskip}
\end{tabular}
\end{center}
\end{table}

   Determinations of the relative distance between the Fornax and
Virgo clusters are quite insensitive to selection effects and
differences of population size, because one compares more or less
comparable galaxies. Therefore the surface brightness fluctuation
(SBF) and planetary nebulae (PNe) methods have been included here
although they provide in general very unreliable distances. It is
known that the bright tail of the luminosity function of the shells of
PNe is strongly dependent on the sample (galaxy) size
\citep{Bottinelli:etal:91,Tammann:93,Mendez:etal:93,Soffner:etal:96}.
The reasons for the generally poor results from SBFs are less
clear. Trouble-makers could be, for instance, AGB stars of an
admixture of a young population or some very metal-rich stars
\citep[cf.][]{Han:etal:97}. 
In any case, one of the strong objections against
the SBF distances is that twelve such distances of galaxies which have
provided blue SNe\,Ia imply a SNe\,Ia luminosity fully $0\mag5$ {\em
  fainter\/} than required by the eight Cepheid-calibrated SNe\,Ia.

   Combining the evidence of the Fornax SNe\,Ia and Table~\ref{tab:9}
yields a weighted distance modulus of the Fornax cluster of
$(m-M)=31.82\pm0.09$. 

   The disagreement of $\Delta(m-M)=0.47\pm0.13$ between the distance
moduli from GCLF and from independent evidence is blatant. It is clear
that in the case of the Fornax cluster the GCLF yields a fallacious
distance. The peak of the GCLF of Fornax must be brighter by
$\sim\!0\mag5$ than in the cases so far considered.

   It should be stressed that all distance determinations of the
Fornax cluster considered here involve early-type galaxies (the
GCLF, SNe\,Ia and Table~\ref{tab:9}). The relatively small Cepheid
distance of the giant {\em spiral\/} NGC\,1365 \citep{Madore:etal:98}
should therefore not be considered here. This is even more the case as
the early-type Fornax galaxies are embedded in a large halo of spiral
galaxies whose mean distance seems somewhat smaller than that of the
E/S0 galaxies \citep{Tammann:Federspiel:97}.

   It may be noted in passing that the distance of the Fornax cluster
is important in the present context to test the GCLF as a distance
indicator, but it has little relevance for the derivation of $H_0$
because the cluster lies well outside the Virgo complex; its peculiar
streaming velocity may therefore be $200\!-\!300\kms$, i.e. 20\% of
the observed mean recession velocity.

\subsection{Individual galaxies}
\label{sec:3_5}

   For six galaxies outside of the Leo group and the big clusters
$m^{\ast}_{\rm V}$ values are in the literature. They are compiled
in Table~\ref{tab:10}.

%
\begin{table}[ht]
\tablefontsize
\begin{center}
\caption{$m^{\ast}_{\rm V}$ values for individual galaxies}
\label{tab:10}
\begin{minipage}{0.9\textwidth}
\begin{tabular}{lrlrrr}
\noalign{\smallskip}
\hline
\noalign{\smallskip}
  Galaxy &  $m^{\ast}_{\rm V}$ & Source & $(m-M)_{\rm GCLF}$ & 
  $v_{220}^{4)}$ & $H_0$ \\
\noalign{\smallskip}
\hline
\noalign{\smallskip}
 NGC\,4278$^{1)}$  & $23.23\pm0.11$ & (1)     & $30.85\pm0.14$ &  542 & 37 \\ 
 NGC\,4494$^{1)}$  & $23.23\pm0.12$ & (1)(2)  & $30.85\pm0.14$ & 1745 & 118 \\ 
 NGC\,4565 (Sb!)$^{1)}$ & $22.63\pm0.20$ & (2) & $30.35\pm0.22$ & 1662 & 141 \\ 
 NGC\,1407$^{2)}$  & $23.95\pm0.30$ & (3)     & $31.57\pm0.31$ & 1710 & 83 \\ 
 NGC\,1400$^{2)}$  & $24.75\pm0.30$ & (3)     & $32.37\pm0.31$ &  475 & 16 \\ 
 NGC\,5846$^{3)}$  & $25.05\pm0.10$ & (4)     & $32.67\pm0.13$ & 1993 & 58 \\
 NGC\,3115         & $22.37\pm0.05$ & (5)     & $29.99\pm0.10$ &  483 & 49 \\
\noalign{\smallskip}
\hline
\noalign{\smallskip}
\end{tabular}
\mpspacing{0.8}

{\footnotesize
 $^{1)}$ Coma\,I cloud (=G13; \citealt{deVaucouleurs:75});
 $^{2)}$ Eridanus cloud (=G31; \citealt{deVaucouleurs:75} does
 not include NGC\,1400);
 $^{3)}$ NGC\,5846 group (=G50; \citealt{deVaucouleurs:60});
 $^{4)}$ Velocity corrected for Virgocentric flow \citep{Kraan-Korteweg:86} 

  Sources: 
 (1) \citealt{Forbes:96b} 
 (2) \citealt{Fleming:etal:95}
 (3) \citealt{Perrett:etal:97}
 (4) \citealt*{Forbes:etal:96}
 (5) \citealt{Kundu:Whitmore:98}.
}
\end{minipage}
\end{center}
\end{table}

   No useful independent distance information other than the
recession velocities is available for the galaxies in
Table~\ref{tab:10}, the exception being NGC\,3115 for which the tip of
the red-giant branch gives a modulus of $(m-M)=30.21\pm0.3$
\citep{Elson:97} in statistical agreement with the GCLF distance.
The individual Hubble constants $H_0$ in the least column of Table~\ref{tab:10}
resulting from the GCLF distances, as determined from $m_{\rm
  V}^{\ast}$ and equation~(\ref{equ:7}), scatter so wildly that the
distances are clearly erratic. To be consistent NGC\,4565 would have
to have a peculiar velocity of $\sim\!1000\kms$, unparalleled outside
of clusters. Also the velocities of NGC\,4278 and NGC\,4494, having
about equal GCLF distances, differ by $>\!1000\kms$. NGC\,1400 is most
likely in the foreground of the Eridanus cloud, yet its GCLF distance
is $0\mag8$ {\em larger\/} than that of the cloud member
NGC\,1407. The conclusion is that the distances in Table~\ref{tab:10},
with the exception of NGC\,5846 and NGC\,3115 and possibly NGC\,4278
and NGC\,1407, are unrealistic. While the peak of the GCLF of
NGC\,4494 and NGC\,4565 seems to be overluminous by $\sim\!1\mag5$ it
is underluminous in NGC\,1400 by almost 3 magnitudes!
These numbers may be somewhat reduced by allowing for peculiar
motions, but it is improbable that the discrepancies will disappear.

\section{Conclusions}
\label{sec:conclusions}

The distances derived from the peak of the GCLF are of highly variable
quality. While the method yields excellent distances to the Leo group,
Virgo cluster, and also the Coma cluster, the distance of the E/S0
galaxies in the Fornax cluster is off by $\sim\!0\mag5$. Even larger
discrepancies are found for at least three individual galaxies in
clouds. The GCLF peak seems to be brighter than average in Fornax,
NGC\,4494, and NGC\,4565 and much fainter in NGC\,1400. It is also
somewhat worrisome that the particularly well determined value $m_{\rm
  V}^{\ast}$ of the enormous GC population of the giant elliptical
M\,87 (NGC\,4486) is brighter by $0\mag35\pm0\mag13$ than the mean of
six other Virgo cluster members.

   There is hence strong evidence that $M^{\ast}$ is variable, but it
is by no means clear which parameters govern this variability.

   One can exclude the Hubble type as the principal parameter because
the calibration from spirals (Galaxy, M\,31) provides perfect distances
for the E/S0 galaxies in the Leo group and in the Virgo and Coma
clusters. Inversely, the GCLF distance of the spiral NGC\,4565 is
inacceptable.

   The available evidence concerning the width $\sigma_{\rm M}$ of the GCLF is
rather inconsistent. Some authors force an adopted $\sigma_{\rm M}$ on
the (partial) observations to determine $m^{\ast}$. If the LF is
observed well beyond the peak a simultaneous solution for $m^{\ast}$
and $\sigma_{\rm M}$ is possible; in other cases $m^{\ast}$ and
$\sigma_{\rm M}$ are correlated. But real differences of $\sigma_{\rm
  M}$ are unquestionable. One may compare $\sigma_{\rm M}=1.02$ for
the Galaxy and M\,31 combined versus the well determined value
$\sigma_{\rm M}=1.40$ for NGC\,4486 \citep{Whitmore:etal:95}; but other large
E galaxies have small $\sigma_{\rm M}$, i.e. $\sigma_{\rm
  M}\approx1.1$ for NGC\,4278 and NGC\,4494 \citep{Forbes:96b}.
Yet at present there are too few {\em reliable\/} $\sigma_{\rm M}$
values to correlate them with the distance deviations.

   Unfortunately very little color information is available for the
GCLF peak. For the Galaxy and M\,31 $m_{\rm B}^{\ast} - m_{\rm
  V}^{\ast}=0.69$ is well determined. The peak color of NGC\,4486,
taken at face value, is much {\em redder\/} and its 
$m_{\rm  V}^{\ast}$ is unusually bright compared with other Virgo
galaxies (cf. Table~\ref{tab:4}). Yet the overluminosity of the Fornax
GCLF peak seems to go with a rather {\em blue\/} peak
(cf. Table~\ref{tab:8}). 

   There are almost certainly metallicity differences, as indicated by
color variations in $(C-T1)$ and $(V-I)$, between the GC systems of
different galaxies. The GCs in spirals are more metal-poor than in
ellipticals \citep{Ashman:Zepf:98}, but as noted before that does not
impair the distance determination of the Leo, Virgo, and Coma
ellipticals by means of the calibration in spirals (Galaxy and
M\,31), although stellar population models suggest that GCs of equal
mass are about $0\mag2$ brighter in ellipticals than in their
metal-poor counterparts in spirals \citep*{Ashman:etal:95}. The likely
merger galaxies NGC\,4486
\citep{Whitmore:etal:95,Elson:Santiago:96a,Elson:Santiago:96b}  
and NGC\,5846 \citep{Forbes:96b} have bimodal distributions in
$(V-I)$, but so do many other galaxies, e.g. NGC\,3115
\citep{Kundu:Whitmore:98}. The bimodality is clearly reflected in the
GCLF in $V$ of NGC\,4486 \citep{Elson:Santiago:96b} and probably of
NGC\,3115 \citep{Kundu:Whitmore:98}, but not obviously so in NGC\,5846
\citep{Forbes:etal:96}. 

   No clear picture emerges which second parameter governs the peak
luminosity $m^{\ast}$ of the GCLF, but in view of the many differences
of GC systems and their apparently different formation histories and
ages it is not too surprising that $m^{\ast}$ yields
distances which are not reliable in all cases.

\acknowledgments
Support of the Swiss National Science Foundation
is gratefully acknowledged.\ The authors thank the organizers,
Dres.~D.~Egret and A.~Heck, for a most stimulating Conference.\
They also thank Mr.~Bernd Reindl for computational help and the outlay
of the manuscript.

%

\begin{thebibliography}{86}
\newcommand{\enquote}[1]{`#1'}
\expandafter\ifx\csname natexlab\endcsname\relax\def\natexlab#1{#1}\fi

\bibitem[{Ajhar et~al.(1994)Ajhar, Blakeslee, \& Tonry}]{Ajhar:etal:94}
Ajhar, E.~A., Blakeslee, J.~P., \& Tonry, J.~L. 1994, AJ, 108, 2087

\bibitem[{Ashman et~al.(1995)Ashman, Conti, \& Zepf}]{Ashman:etal:95}
Ashman, K.~M., Conti, A., \& Zepf, S.~E. 1995, AJ, 110, 1164

\bibitem[{Ashman \& Zepf(1998)}]{Ashman:Zepf:98}
Ashman, K.~M. \& Zepf, S.~E. 1998, in \enquote{Globular Cluster Systems},
  Cambridge: Cambridge Univ. Press, p.~89

\bibitem[{Baum et~al.(1995)Baum, Hammergren, Groth, Ajhar, \&
  Lauer}]{Baum:etal:95}
Baum, W.~A., Hammergren, M., Groth, E.~J., Ajhar, E.~A., \& Lauer, T.~R. 1995,
  AJ, 110, 2537

\bibitem[{Baum et~al.(1997)Baum, Hammergren, Thomsen, Groth, Faber, Grillmair,
  \& Ajhar}]{Baum:etal:97}
Baum, W.~A., Hammergren, M., Thomsen, B., Groth, E.~J., Faber, S.~M.,
  Grillmair, C.~J., \& Ajhar, E.~A. 1997, AJ, 113, 1483

\bibitem[{Binggeli et~al.(1985)Binggeli, Sandage, \&
  Tammann}]{Binggeli:etal:85}
Binggeli, B., Sandage, A., \& Tammann, G.~A. 1985, AJ, 90, 1681

\bibitem[{Blakeslee \& Tonry(1996)}]{Blakeslee:Tonry:96}
Blakeslee, J.~P. \& Tonry, J.~L. 1996, ApJ, 465, L19

\bibitem[{Bottinelli et~al.(1991)Bottinelli, Gougenheim, Paturel, \&
  Teerkorpi}]{Bottinelli:etal:91}
Bottinelli, L., Gougenheim, L., Paturel, G., \& Teerkorpi, P. 1991, A\&A, 252,
  560

\bibitem[{Bridges et~al.(1991)Bridges, Hanes, \& Harris}]{Bridges:etal:91}
Bridges, T.~J., Hanes, D.~A., \& Harris, W.~E. 1991, AJ, 101, 469

\bibitem[{Burstein \& Heiles(1984)}]{Burstein:Heiles:84}
Burstein, D. \& Heiles, C. 1984, ApJS, 54, 33

\bibitem[{Cacciari(1998)}]{Cacciari:98}
Cacciari, C. 1998, private communication

\bibitem[{Capaccioli et~al.(1989)Capaccioli, Della~Valle, Rosino, \&
  D'Onofrio}]{Capaccioli:etal:89}
Capaccioli, M., Della~Valle, M., Rosino, L., \& D'Onofrio, M. 1989, AJ, 97,
  1622

\bibitem[{Chaboyer et~al.(1998)Chaboyer, Demarque, Kernan, \&
  Krauss}]{Chaboyer:etal:98}
Chaboyer, B., Demarque, P., Kernan, P.~J., \& Krauss, L.~M. 1998, ApJ, 494, 96

\bibitem[{Cohen(1988)}]{Cohen:88}
Cohen, J. 1988, AJ, 95, 682

\bibitem[{Dekel(1995)}]{Dekel:95}
Dekel, A. 1995, private communication

\bibitem[{{Della Valle} et~al.(1998){Della Valle}, Kissler-Patig, Danziger, \&
  Storm}]{DellaValle:etal:98}
{Della Valle}, M., Kissler-Patig, M., Danziger, J., \& Storm, J. 1998, MNRAS,
  299, 267

\bibitem[{{de Vaucouleurs}(1960)}]{deVaucouleurs:60}
{de Vaucouleurs}, G. 1960, ApJ, 131, 585

\bibitem[{{de Vaucouleurs}(1970)}]{deVaucouleurs:70}
{de Vaucouleurs}, G. 1970, ApJ, 159, 435

\bibitem[{{de Vaucouleurs}(1975)}]{deVaucouleurs:75}
{de Vaucouleurs}, G. 1975, in \enquote{Galaxies and the Universe}, eds.
  A.~Sandage \& J.~Kristian, Chicago: Univ. of Chicago Press, p. 557

\bibitem[{D'Onofrio et~al.(1997)D'Onofrio, Capaccioli, Zaggia, \&
  Caon}]{DOnofrio:etal:97}
D'Onofrio, M., Capaccioli, M., Zaggia, S.~R., \& Caon, N. 1997, MNRAS, 289, 847

\bibitem[{Dressler(1987)}]{Dressler:87}
Dressler, A. 1987, ApJ, 317, 1

\bibitem[{Elson(1997)}]{Elson:97}
Elson, R. A.~W. 1997, MNRAS, 286, 771

\bibitem[{Elson \& Santiago(1996{\natexlab{a}})}]{Elson:Santiago:96a}
Elson, R. A.~W. \& Santiago, B.~X. 1996{\natexlab{a}}, MNRAS, 278, 617

\bibitem[{Elson \& Santiago(1996{\natexlab{b}})}]{Elson:Santiago:96b}
Elson, R. A.~W. \& Santiago, B.~X. 1996{\natexlab{b}}, MNRAS, 280, 971

\bibitem[{Faber et~al.(1989)Faber, Wegner, Burstein, Davies, Dressler,
  Lynden-Bell, \& Terlevich}]{Faber:etal:89}
Faber, S.~M., Wegner, G., Burstein, D., Davies, R.~L., Dressler, A.,
  Lynden-Bell, D., \& Terlevich, R.~J. 1989, ApJS, 69, 763

\bibitem[{Federspiel et~al.(1998)Federspiel, Tammann, \&
  Sandage}]{Federspiel:etal:98}
Federspiel, M., Tammann, G.~A., \& Sandage, A. 1998, ApJ, 495, 115

\bibitem[{Fernley et~al.(1998)}]{Fernley:etal:98}
Fernley, J. et~al. 1998, this conference

\bibitem[{Fleming et~al.(1995)Fleming, Harris, Pritchet, \&
  Hanes}]{Fleming:etal:95}
Fleming, D. E.~B., Harris, W.~E., Pritchet, C.~J., \& Hanes, D. 1995, AJ, 109,
  1044

\bibitem[{Forbes(1996{\natexlab{a}})}]{Forbes:96a}
Forbes, D.~A. 1996{\natexlab{a}}, AJ, 112, 954

\bibitem[{Forbes(1996{\natexlab{b}})}]{Forbes:96b}
Forbes, D.~A. 1996{\natexlab{b}}, AJ, 112, 1409

\bibitem[{Forbes et~al.(1996)Forbes, Brodie, \& Huchra}]{Forbes:etal:96}
Forbes, D.~A., Brodie, J.~P., \& Huchra, J. 1996, AJ, 112, 2448

\bibitem[{Geisler \& Forte(1990)}]{Geisler:Forte:90}
Geisler, D. \& Forte, J.~C. 1990, ApJ, 350, L5

\bibitem[{Graham et~al.(1997)Graham, et~al.}]{Graham:etal:97}
Graham, J.~A., et~al. 1997, ApJ, 477, 535

\bibitem[{Gratton et~al.(1997)Gratton, {Fusi Pecci}, Carretta, Clementini,
  Corsi, \& Lattanzi}]{Gratton:etal:97}
Gratton, R.~G., {Fusi Pecci}, F., Carretta, E., Clementini, G., Corsi, C.~E.,
  \& Lattanzi, M. 1997, ApJ, 491, 749

\bibitem[{Han et~al.(1997)}]{Han:etal:97}
Han, M. et~al. 1997, AJ, 113, 1001

\bibitem[{Harris(1990)}]{Harris:90}
Harris, W.~E. 1990, PASP, 102, 966

\bibitem[{Harris(1991)}]{Harris:91}
Harris, W.~E. 1991, ARA\&A, 29, 543

\bibitem[{Harris et~al.(1991)Harris, Allwright, Pritchet, \& {van den
  Bergh}}]{Harris:etal:91}
Harris, W.~E., Allwright, J. W.~B., Pritchet, C.~J., \& {van den Bergh}, S.
  1991, ApJS, 76, 115

\bibitem[{Hawley et~al.(1986)Hawley, Jeffreys, Barnes, \& Lai}]{Hawley:etal:86}
Hawley, S.~L., Jeffreys, W.~H., Barnes, T.~G., \& Lai, W. 1986, ApJ, 302, 626

\bibitem[{Holland(1998)}]{Holland:98}
Holland, S. 1998, AJ, 115, 1916

\bibitem[{Hubble(1932)}]{Hubble:32}
Hubble, E. 1932, ApJ, 76, 44

\bibitem[{Humason et~al.(1956)Humason, Mayall, \& Sandage}]{Humason:etal:56}
Humason, M.~L., Mayall, N.~U., \& Sandage, A.~R. 1956, AJ, 61, 97

\bibitem[{Jacoby(1997)}]{Jacoby:97}
Jacoby, G.~H. 1997, in \enquote{The Extragalactic Distance Scale}, eds.
  M.~Livio, M.~Donahue, \& N.~Panagia, Cambridge: Cambridge Univ. Press, p. 186

\bibitem[{Jerjen \& Binggeli(1997)}]{Jerjen:Binggeli:97}
Jerjen, H. \& Binggeli, B. 1997, in \enquote{The Nature of Elliptical
  Galaxies},  ASP Conference Series 116, p. 298

\bibitem[{Kissler et~al.(1994)Kissler, Richtler, Held, Grebel, Wagner, \&
  Capaccioli}]{Kissler:etal:94}
Kissler, M., Richtler, T., Held, E.~V., Grebel, E.~K., Wagner, S.~J., \&
  Capaccioli, M. 1994, A\&A, 287, 463

\bibitem[{Kohle et~al.(1996)Kohle, Kissler-Patig, Hilker, Richtler, Infante, \&
  Quintana}]{Kohle:etal:96}
Kohle, S., Kissler-Patig, M., Hilker, M., Richtler, T., Infante, L., \&
  Quintana, H. 1996, A\&A, 309, L39

\bibitem[{Kraan-Korteweg(1986)}]{Kraan-Korteweg:86}
Kraan-Korteweg, R.~C. 1986, A\&AS, 66, 255

\bibitem[{Kundu \& Whitmore(1998)}]{Kundu:Whitmore:98}
Kundu, A. \& Whitmore, B.~C. 1998, preprint

\bibitem[{Livio(1997)}]{Livio:97}
Livio, M. 1997, in \enquote{The Extragalactic Distance Scale}, eds. M.~Livio,
  M.~Donahue, \& N.~Panagia, Cambridge: Cambridge Univ. Press, p. 186

\bibitem[{Macchetto(1998)}]{Macchetto:98}
Macchetto, F.~D. 1998, this conference

\bibitem[{Madejsky \& Bender(1990)}]{Madejsky:Bender:90}
Madejsky, R. \& Bender, R. 1990, IAU Symp., 139, 377

\bibitem[{Madore \& Freedman(1991)}]{Madore:Freedman:91}
Madore, B. \& Freedman, W.~L. 1991, PASP, 103, 933

\bibitem[{Madore et~al.(1998)}]{Madore:etal:98}
Madore, B.~F. et~al. 1998, Nature, 395, 47

\bibitem[{McLaughlin et~al.(1994)McLaughlin, Harris, \&
  Hanes}]{McLaughlin:etal:94}
McLaughlin, D.~E., Harris, W.~E., \& Hanes, D.~A. 1994, ApJ, 422, 486

\bibitem[{M{\'e}ndez et~al.(1993)M{\'e}ndez, Kudritzki, Ciardullo, \&
  Jacoby}]{Mendez:etal:93}
M{\'e}ndez, R.~H., Kudritzki, R.~P., Ciardullo, R., \& Jacoby, G.~H. 1993,
  A\&A, 275, 534

\bibitem[{Mould \& Kristian(1986)}]{Mould:Kristian:86}
Mould, J. \& Kristian, J. 1986, ApJ, 305, 591

\bibitem[{Parodi et~al.(1998)Parodi, Saha, Sandage, \&
  Tammann}]{Parodi:etal:98}
Parodi, B., Saha, A., Sandage, A., \& Tammann, G.~A. 1998, preprint

\bibitem[{Perrett et~al.(1997)Perrett, Hanes, Butterworth, \&
  Kavelaars}]{Perrett:etal:97}
Perrett, K.~M., Hanes, D.~A., Butterworth, S.~T., \& Kavelaars, J.~J. 1997, AJ,
  113, 895

\bibitem[{Pritchet \& {van den Bergh}(1985)}]{Pritchet:vandenBergh:85}
Pritchet, C.~J. \& {van den Bergh}, S. 1985, AJ, 90, 2027

\bibitem[{Pritchet \& {van den
  Bergh}(1987{\natexlab{a}})}]{Pritchet:vandenBergh:87a}
Pritchet, C.~J. \& {van den Bergh}, S. 1987{\natexlab{a}}, ApJ, 316, 517

\bibitem[{Pritchet \& {van den
  Bergh}(1987{\natexlab{b}})}]{Pritchet:vandenBergh:87}
Pritchet, C.~J. \& {van den Bergh}, S. 1987{\natexlab{b}}, ApJ, 318, 507

\bibitem[{Racine(1968)}]{Racine:68}
Racine, R. 1968, JRASC, 62, 367

\bibitem[{Reid(1998)}]{Reid:98}
Reid, I.~N. 1998, AJ, 115, 204

\bibitem[{Richtler et~al.(1992)}]{Richtler:etal:92}
Richtler, T. et~al. 1992, A\&A, 264, 25

\bibitem[{Saha et~al.(1996)Saha, Sandage, Labhardt, Tammann, Macchetto, \&
  Panagia}]{Saha:etal:96a}
Saha, A., Sandage, A., Labhardt, L., Tammann, G.~A., Macchetto, F.~D., \&
  Panagia, N. 1996, ApJ, 466, 55

\bibitem[{Saha et~al.(1998)Saha, Sandage, Labhardt, Tammann, Macchetto, \&
  Panagia}]{Saha:etal:98a}
Saha, A., Sandage, A., Labhardt, L., Tammann, G.~A., Macchetto, F.~D., \&
  Panagia, N. 1998, preprint

\bibitem[{Sandage(1968)}]{Sandage:68}
Sandage, A. 1968, ApJl, 152, L149

\bibitem[{Sandage(1993)}]{Sandage:93c}
Sandage, A. 1993, AJ, 106, 703

\bibitem[{Sandage \& Bedke(1988)}]{Sandage:Bedke:88}
Sandage, A. \& Bedke, J. 1988, Atlas of Galaxies useful to measure the
  Cosmological Distance Scale, Washington: NASA

\bibitem[{Sandage et~al.(1976)Sandage, Kristian, \& Westphal}]{Sandage:etal:76}
Sandage, A., Kristian, J., \& Westphal, J.~A. 1976, ApJ, 205, 688

\bibitem[{Sandage \& Tammann(1988)}]{Sandage:Tammann:88}
Sandage, A. \& Tammann, G.~A. 1988, ApJ, 328, 1

\bibitem[{Sandage \& Tammann(1995)}]{Sandage:Tammann:95}
Sandage, A. \& Tammann, G.~A. 1995, ApJ, 446, 1

\bibitem[{Sandage \& Tammann(1996)}]{Sandage:Tammann:96a}
Sandage, A. \& Tammann, G.~A. 1996, ApJ, 464, L51

\bibitem[{Secker(1992)}]{Secker:92}
Secker, J. 1992, AJ, 104, 1472

\bibitem[{Secker \& Harris(1993)}]{Secker:Harris:93}
Secker, J. \& Harris, W.~E. 1993, AJ, 105, 1358

\bibitem[{Soffner et~al.(1996)Soffner, M{\'e}ndez, Jacoby, Ciardullo, Roth, \&
  Kudritzki}]{Soffner:etal:96}
Soffner, T., M{\'e}ndez, R., Jacoby, G., Ciardullo, R., Roth, M., \& Kudritzki,
  R. 1996, A\&A, 306, 9

\bibitem[{Tammann(1988)}]{Tammann:88}
Tammann, G.~A. 1988, in \enquote{The Extragalactic Distance Scale}, eds.
  S.~van~den Bergh \& C.~J. Pritchet, San Francisco: ASP, p. 282

\bibitem[{Tammann(1993)}]{Tammann:93}
Tammann, G.~A. 1993, in \enquote{Planetary Nebulae}, eds. R.~Weinberger \&
  A.~Acker, Dordrecht: Kluwer,  IAU Symp. 155, p. 515

\bibitem[{Tammann \& Federspiel(1997)}]{Tammann:Federspiel:97}
Tammann, G.~A. \& Federspiel, M. 1997, in \enquote{The Extragalactic Distance
  Scale}, eds. M.~Livio, M.~Donahue, \& N.~Panagia, Cambridge: Cambridge Univ.
  Press, p. 137

\bibitem[{Tanvir et~al.(1995)Tanvir, Shanks, Ferguson, \&
  Robinson}]{Tanvir:etal:95}
Tanvir, N.~R., Shanks, T., Ferguson, H.~C., \& Robinson, D. R.~T. 1995, Nature,
  377, 27

\bibitem[{Tonry(1997)}]{Tonry:97}
Tonry, J.~L. 1997, in \enquote{The Extragalactic Distance Scale}, eds.
  M.~Livio, M.~Donahue, \& N.~Panagia, Cambridge: Cambridge Univ. Press, p. 297

\bibitem[{Tsujimoto(1998)}]{Tsujimoto:98}
Tsujimoto, T. 1998, this conference

\bibitem[{{van den Bergh} et~al.(1985){van den Bergh}, Pritchet, \&
  Grillmair}]{vandenBergh:etal:85}
{van den Bergh}, S., Pritchet, C., \& Grillmair, C. 1985, AJ, 90, 595

\bibitem[{Weedman(1976)}]{Weedman:76}
Weedman, D.~W. 1976, ApJ, 203, 6

\bibitem[{Whitmore(1997)}]{Whitmore:97}
Whitmore, B.~C. 1997, in \enquote{The Extragalactic Distance Scale}, eds.
  M.~Livio, M.~Donahue, \& N.~Panagia, Cambridge: Cambridge Univ. Press, p. 254

\bibitem[{Whitmore et~al.(1995)Whitmore, Sparks, Lucas, Macchetto, \&
  Biretta}]{Whitmore:etal:95}
Whitmore, B.~C., Sparks, W.~B., Lucas, R.~A., Macchetto, F.~D., \& Biretta,
  J.~A. 1995, ApJ, 454, L73

\end{thebibliography}

%
\end{document}